\documentclass[letterpaper, 10 pt, conference]{ieeeconf}  

\IEEEoverridecommandlockouts 
\usepackage{graphics} 
\usepackage{epsfig} 
\usepackage{mathptmx} 
\usepackage{times} 
\usepackage{amsmath} 
\usepackage{amssymb}  

\usepackage[caption=false]{subfig}

\usepackage{xcolor}
\usepackage[normalem]{ulem}
\definecolor{correction}{RGB}{255, 0, 0}

\definecolor{black}{RGB}{0, 0, 0}
\definecolor{orange}{RGB}{230, 159, 0}
\definecolor{skyblue}{RGB}{86, 180, 233}
\definecolor{green}{RGB}{0, 158, 115}
\definecolor{yellow}{RGB}{240, 228, 66}
\definecolor{blue}{RGB}{0, 114, 178}
\definecolor{red}{RGB}{213, 94, 0}
\definecolor{purple}{RGB}{204, 121, 167}
\definecolor{light-gray}{gray}{0.70}
\definecolor{darkgray176}{RGB}{176,176,176}
\definecolor{darkkhaki}{RGB}{189,183,107}
\definecolor{darkred}{RGB}{139,0,0}
\definecolor{dimgray}{RGB}{105,105,105}
\definecolor{gray}{RGB}{128,128,128}
\definecolor{navy}{RGB}{0,0,128}
\usepackage[utf8]{inputenc}
\usepackage{pgfplots}
\DeclareUnicodeCharacter{2212}{−}
\usepgfplotslibrary{groupplots,dateplot}
\usepackage{tikz}
\usetikzlibrary{shapes.geometric}
\usetikzlibrary{arrows.meta,arrows}
\usetikzlibrary{patterns,shapes.arrows}
\pgfplotsset{compat=newest}
\usepackage{environ}
\makeatletter
\newsavebox{\measure@tikzpicture}
\NewEnviron{scaletikzpicturetowidth}[1]{%
  \def\tikz@width{#1}%
  \begin{lrbox}{\measure@tikzpicture}%
  \BODY
  \end{lrbox}%
  \pgfmathparse{#1/\wd\measure@tikzpicture}%
  \BODY
}
\makeatother

\usepackage{multirow}

\usepackage{hyperref}       
\definecolor{dark-red}{rgb}{0.4,0.15,0.15}
\definecolor{dark-blue}{rgb}{0.15,0.15,0.8}
\definecolor{medium-blue}{rgb}{0,0,0.5}
\hypersetup{
    colorlinks, linkcolor={dark-red},
    citecolor={dark-blue}, urlcolor={medium-blue}
}

\title{\LARGE \bf
Automatic Classification of Subjective Time Perception Using Multi-modal Physiological Data of Air Traffic Controllers
}

\author{Till Aust,$^{1}$ Eirini Balta,$^{2}$ Argiro Vatakis,$^{2}$ and Heiko Hamann$^{1}$
\thanks{$^{1}$Till Aust and Heiko Hamann are with the Department of Computer and Information Science, University of Konstanz, Konstanz, Germany. 
{\tt\small till.aust@uni-konstanz.de}}%
\thanks{$^{2}$Eirini Balta and Argiro Vatakis are with the Department of Psychology, Panteion University of Social and Political Sciences, Athens, Greece. 
{\tt\small argiro.vatakis@panteion.gr}}%
}

\begin{document}

\maketitle
\thispagestyle{empty}
\pagestyle{empty}

\begin{abstract} 
In high-pressure environments where human individuals must simultaneously monitor multiple entities, communicate effectively, and maintain intense focus, the perception of time becomes a critical factor influencing performance and well-being.  
One indicator of well-being can be the person's subjective time perception. 
In our project \textit{ChronoPilot}, we aim to develop a device that modulates human subjective time perception. 
In this study, we present a method to automatically assess the subjective time perception of air traffic controllers, a group often faced with demanding conditions, using their physiological data and eleven state-of-the-art machine learning classifiers. 
The physiological data consist of photoplethysmogram, electrodermal activity, and temperature data. 
We find that the support vector classifier works best with an accuracy of 79~\% and electrodermal activity provides the most descriptive biomarker. 
These findings are an important step towards closing the feedback loop of our \textit{ChronoPilot}-device to automatically modulate the user's subjective time perception. 
This technological advancement may promise improvements in task management, stress reduction, and overall productivity in high-stakes professions.
\end{abstract}






\section{Introduction} 
\label{sec:introduction}
In a rapidly evolving world, prioritizing mental health becomes more crucial especially among the working population~\cite{McGrath2023}. 
Work can be a significant stressor, which may reduce well-being, lower productivity, or even lead to burn out. 
One indicator can be the subjective perceived passage of time, which then passes slower~\cite{Ogden2020}. 
When working individuals are positively engaged they perceive time faster. 

\begin{figure}
    \centering
    \hspace*{-0.2cm}
    \input{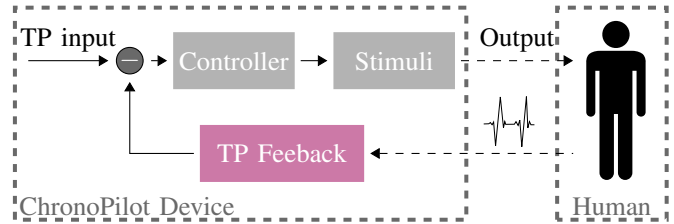}
    \caption{Fundamental control loop of the envisioned \textit{ChronoPilot}-device. \textit{TP Input} is the user's desired time perception for maximal productivity and well-being. The \textit{Controller} produces \textit{Stimuli} and applies them to the user. Physiological data are collected as feedback for the \textit{ChronoPilot}-device to adjust the stimuli accordingly. Our objective is to develop machine learning methods, which are able to provide this feedback by classifying the perceived passage of time based on physiological data.}
    \label{fig:feedback_loop}
\end{figure}

In our EU-funded project \textit{ChronoPilot}~\cite{Botev2021}, we are developing a device that modulates human time perception to increase performance and well-being of human workers. 
For this so-called \textit{ChronoPilot}-device, we propose a control loop (see Fig.~\ref{fig:feedback_loop}), which interacts with the human using haptic~\cite{Cavdan2023}, visual~\cite{Schatzschneider2016}, auditory~\cite{Picard2022}, or situational stimuli~\cite{Kaduk2023}, and gathers non-invasive physiological data. 
Our objective in this paper is to present a methodology to extract automated feedback on the user's subjective perceived passage of time out of the physiological data for the \textit{ChronoPilot}-device. 
To do so, we derive meaningful features from the physiological data, also called biomarkers, for classification, similar to related fields of automated detection of stress~\cite{Gedam2020} or emotion~\cite{Kumar2022}.   

A common approach to obtain automated feedback from humans is to collect their physiological data using wearable sensing devices~\cite{Montesinos2019,Orlandic2021,Schmidt2019}. 
Montesinos~\textit{et al.}~\cite{Montesinos2019} used wearable devices to implement multi-modal acute stress recognition. 
We follow a similar pipeline as proposed by them, consisting of signal acquisition, data processing, and model learning and evaluation.
Unlike other works in this domain, they tested on unseen data to validate the generalizability of their trained classifiers. 
We propose to use leave-one-subject-out cross-validation (validating our classifiers on data not used for training) to prove generalizability.
Masinelli~\textit{et al.}~\cite{Masinelli2020} use wearable sensors in combination with machine learning methods to monitor cognitive load of individual users. 
We advance this study by classifying for subjective perceived passage of time. 

The only study we are aware of that is classifying subjective time perception based on physiological signals was done by Orlandic~\textit{et al.}~\cite{Orlandic2021}. 
They stimulated users with different video clips selected from the Emotional Movie~\cite{Carvalho2012} and FilmStim~\cite{Schaefer2010} databases, which lead to different emotional states (e.g., neutral, fear, sadness) along with cognitive tasks, such as mathematical tasks or the Stroop Color Test~(varies cognitive load through congruent and incongruent stimuli).
They collected skin temperature, electrodermal activity, respiration, electrocardiogram, and photoplethysmogram data. 
The feature extraction was done on non-overlapping 45-second slices, which were normalized by subtracting the mean and dividing by the standard deviation of the training dataset. 
They then applied a set of machine learning classification algorithms. 
In contrast, we use data of a more complex real-world scenario, namely flight control, instead of an in-lab experimental setup. 
We assess a different aspect of subjective time, that is, the subjective perceived passage of time (time passage) instead of the relative estimation error based on verbal time estimation (duration). 
Our two main contributions are: (1)~an automatic multi-modal subjective perceived passage of time perception estimation technique, using photoplethysmogram, electrodermal activity, and temperature data measured by of-the-shelf wearable sensors, with which, (2)~we obtain high accuracies of up to 79~\% on real-world physiological data obtained during flight controller helicopter training sequences. 
Our analysis includes an evaluation and discussion of eleven state-of-the-art machine learning classifiers for subjective perceived passage of time classification in combination with feature selection methods and feature importance analysis using the \textit{SHapley Additive exPlanations}~(SHAP)~\cite{Lundberg2017} framework. 
We show generalizability by validating our methods using leave-one-subject-out cross validation~(LOSOCV).

\section{Scenario and Methodology}
\label{sec:methodology}
We first describe the air traffic control scenario under various workload manipulations in real-world conditions~\cite{Balta2024}. 
We present our overall methodology consisting of data acquisition, preprocessing, and classification. 
Preprocessing includes manual cleaning, filtering, segmentation, background subtraction, and feature extraction. 
The classification consists of machine learning classifiers and feature selection methods.

\subsection{Scenario}
\label{sec:scenario}
Twelve air traffic controllers (4 females; mean age = 36.9, SD = 6.6), participated in the experiment. 
The participants were all professionals, certified by the Civil Aviation Authority of the Ministry of Infrastructure and Transport (Greece), and working at a Greek military airport. 
Their native language was Greek, they all had an excellent knowledge of English as a second language, and they had undergone the necessary training in air traffic management terminology.
In all experimental conditions, the air-traffic controllers were wearing the EmotiBit device\footnote{\url{https://www.emotibit.com}} and completed a flying control session, consisting of a scheduled training flight for pilots of military helicopters, where the pilots had to take-off, perform a full circle over the airport, and land. 
During the cruising phase the controllers only maintained visual contact with the helicopter, while during the landing phase the controllers were actively engaged in the process. 
At the end of each experimental session, each controller made a verbal time estimation of the total session duration. 
Additionally, the controllers gave a rating on how slow or fast they felt the time had passed during the session (i.e., passage of time judgments) on an integer scale from~1~to~5. 
The level of workload was varied by the number of helicopters controlled by the air traffic controller, either~1~or~2, and the language of communication between the controller and the pilot of the helicopter, either Greek or English. 
The experiment took place on two consecutive days, between the hours of 08:00~am and 02:00~pm with similar weather conditions. 
Each air traffic controller participated in two sessions per day, for a total of four sessions. 
The order of the sessions was randomized.

\subsection{Data acquisition}
\label{sec:experiment}
The experiment was conducted in the air traffic control premises located on a military airport. 
The main equipment used was the communication device, and the digital recorder of the control tower. 
The communication device was an Integrated Communications (ICOM) VHF air-band base station radio, equipped with a hand-held microphone that the controllers used to communicate with the pilots. 
The EmotiBit open-source wearable device was used to measure the cardiac and electrodermal activity throughout experimentation. 
The 3-wavelength photoplethysmogram (PPG) sensor was used to monitor blood volume changes at a sampling rate of 25~Hz, and the electrodermal activity (EDA) sensor was used to monitor skin conductance at a sampling rate of 15~Hz. 
The physiological data were recorded on a secure digital card (SD card) for offline processing.
Hence, we have a total of 48 sequences with an average length of 182~s. 
For preprocessing, we sort the sequences per participant and increasing mental load, that is, from one helicopter communicating in Greek to two helicopters communicating in English (non-native language). 

\subsection{Data processing}
\label{sec:preprocessing_steps}

We apply similar preprocessing steps to all obtained physiological signals.\footnote{Code available at: \url{https://github.com/tilly111/chronopilot_feature_extraction}} 
First, we manually clean the signals to remove artifacts caused, for example, by sensor movement. 
Second, we filter the physiological signals to remove measurement noise and split them into the desired segments for background subtraction. 
Finally, we extract psychologically meaningful features from the segments, later used for classification.

\subsubsection{Photoplethysmogram (PPG) data}
Initially, we apply manual cleaning, which consists of removing the last eight and~13 values of sequence 3 from participant~12 and sequence~one of participant~five, which peaked due to artifacts from motion of the measuring device. 
After the manual cleaning, we use the cleaning method from \textit{HeartPy}~\cite{VanGent2018} with cutoff frequencies 0.7~Hz and 3.5~Hz allowing for heart rates from 42~bpm to 210~bpm in the bandpass filter of order~three. 
The original sampling rate is approximately 25~Hz. 
For a more stable feature calculation, we up-sample the signal to approximately 100~Hz using the Fourier method.\footnote{\url{https://docs.scipy.org/doc/scipy/reference/generated/scipy.signal.resample.html}} 
Thereafter, we extract 13~physiological features using the \textit{HeartPy} toolkit for PPG data.

\subsubsection{Electrodermal activity (EDA) data}
Similar to the PPG data, we apply manual cleaning to remove peaks at the end of the same sequences. 
The original sampling rate is approximately 15~Hz. 
For a more stable feature calculation, we up-sampled the signal to approximately $10^4$~Hz using the same Fourier method as for the PPG data. 
From the up-sampled signal, we extract 6~physiological features using the \textit{NeuroKit2} toolkit~\cite{Makowski2021} for EDA data. 
The toolkit automatically cleans the data before extracting the features. 
We extend the signal used for background subtraction of participant~6 in sequence~1 by one and a half seconds to have 10~s, which is the minimum required length for processing the signal.

\subsubsection{Temperature (Thermopile)}
The temperature of the thermopile and one reference surrounding temperature measurement is sampled with a frequency of 7.5~Hz.
We calculate $\delta$ the mean difference between the reference and the thermopile temperature during the whole experiment and the individual means. 
Further, we calculate $\delta$ the gradient of the difference and the total power of its power spectral density (PSD). 
Hence, we extract a total of five features from the temperature data.  
These are similar features to Orlandic \textit{et al.}~\cite{Orlandic2021} with the difference that they measure skin temperature, while we measure the body temperature.

\subsubsection{Normalization}
\label{sec:normalization}
We employed background subtraction to reduce inter-participant variances. 
For the baseline interval, we use the time span between the start of the experiment and the take-off of the helicopter(s). 
We extract the features from the baseline interval to subtract them from the features extracted from the whole experiment. 
On the background subtracted features, we performed min-max-scaling. 
That is, we scaled the data based on the training data between~0~and~1 to eliminate the learning of spurious relations between feature scale and setting. 
This normalization yielded better numerical stability during training. 
We compare the min-max-scaling with z-score normalization and none normalization as a baseline. 
The unseen test data are scaled using the parameters derived from the training data. 

\subsubsection{Labels}
\begin{figure}
     \centering
     \vspace*{0.25cm}
     \input{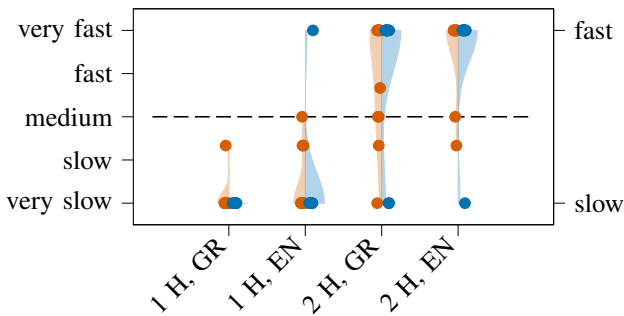}\label{fig:labels_ground_truth}
     \caption{Violin plots of the perceived passage of time labels of the different settings of the flight controller experiment (H:~helicopter number, GR:~Greek, EN:~English; $n=12$). 
     The left side of the violin plot (orange) displays the scaled label distribution from the questionnaire (scale is the left axis). 
     The right side (blue) shows the label distribution after thresholding (dashed line) in the \textit{slow} and \textit{fast} class. 
     The dots represent individual samples (overlapping). 
     We observe that depending on the number of helicopters the majority of the controllers perceived the passage of time as slow when there is a one helicopter and fast when there are two helicopters.}
     \label{fig:label_helicopter}
\end{figure}
For classification, we use the perceived passage of time estimation from the controllers' questionnaires obtained at the end of the experiment. 
To allow for better comparability between questionnaire answers, we min-max-scale each controller's answers from \textit{very slow} to \textit{very fast}, see Fig.~\ref{fig:label_helicopter}, left half (orange). 
We sort the settings by speaking in Greek and one helicopter to speaking in English and two helicopters as we suspect an increase in mental load for speaking in a non-native language and an increased number of to-be-controlled objects. 
This sorting suggest a binary classification problem, namely, \textit{slow} for one helicopter settings and \textit{fast} for two helicopters settings. 
By thresholding the scaled labels at medium ($> 3$ is labeled as fast), we obtain a binary classification problem with 26~sequences of class~\textit{slow} and 22~sequences of class~\textit{fast}, see Fig.~\ref{fig:label_helicopter}, right half (blue). 

We train our models using LOSOCV, which means we train with all but one participant and use this participant for validation (validating on unseen data). 
Hence, we train each classifier in each setting 12~times and average the results. 
We chose this validation strategy given the small dataset and accurately estimating model performance is more important than computational cost~\cite{Luntz1969}.

\subsection{Classification}
\label{sec:classification}

\subsubsection{Classifiers}
For classification, we follow a similar approach as Schmidt~\textit{et~al.}~\cite{Schmidt2019} proposed by combining the extracted features to feature vectors, which we feed into our machine learning classifiers. 
Due to the small dataset, we avoid deep learning classifiers as they are known to require immense amounts of data~\cite{Anthony1999}. 
That is why we use a similar set of state-of-the-art machine learning (ML) classifiers as presented by Orlandic \textit{et al.}~\cite{Orlandic2021}, namely Support Vector Classifier (SVC), Decision-tree Classifier (DTC), k-Nearest Neighbor Classifier (KNN), Logistic Regression (LR), Gaussian Naive Bayes (GNB), Linear Discriminant Analysis (LDA), Random Forest Classifier (RF), and eXtreme Gradient Boosting (XGB). 
In addition, we test Quadratic Discriminant Analysis (QDA), Ada Boost (AB), and Gradient Boosting (GB). 

For most ML classifiers we use the \textit{scikit-learn library}~\cite{scikit-learn} implementation. 
The eXtreme Gradient Boosting is implemented using the \textit{xgboost}\footnote{eXtreme Gradient Boosting classifier \url{https://github.com/dmlc/xgboost}} package. 
The hyperparameter tuning was done manually and the values were chosen individually such that the performance for each classifier was maximized.\footnote{Code available at: \url{https://github.com/tilly111/chronopilot_classification}}

\subsubsection{Feature selecting algorithms}
\label{sec:feature_selecting_algorithms}
We train our classifiers with all features and subsets of features (split by physiological signals). 
Additionally, we test two feature selection algorithms: sequential feature selection (SFS) and recursive feature elimination with cross-validation (RFECV). 
These methods shrink the input space and, thus, should simplify the classification problem and make the classifiers smaller, which improves performance, reduces overfitting, and enhances the interpretability of the model. 

SFS adds (forward selection) or removes (backward selection) features from a feature subset in a greedy fashion. 
The criteria at each stage is that the estimator chooses the best feature to add or remove based on the cross-validation score of the given ML classifier. 

RFECV removes the least important feature at each stage by evaluating the model's performance using cross-validation. 
As the classifier iterates in each stage through all possible features it is not greedy but should yield the best possible feature set. 
Both feature selection algorithms are implemented using the \textit{scikit-learn library}~\cite{scikit-learn}.

\section{Results}
\label{sec:results}
In this paper we work towards providing automated feedback to the envisioned \textit{ChronoPilot}-device based on the physiological data of the human user. 
The device would trigger the corresponding stimuli to modulate the user's time perception. 
The device would operate online, continuously, and individually for each user in real-world settings. 
Hence, we have relaxed requirements for classification accuracy; we could incorporate temporal preceding classification results to boost certainty exploiting the temporal dependency of the problem. 
If the system operates over longer periods of time, we can use online learning techniques to optimize and adapt for the individual user to improve performance even more. 
Despite the ambitious objective of our project, here we only aim to show general feasibility. 
We evaluate a group of users without fine-tuning for individuals. 
Hence, we are satisfied by performance that is better than a randomized approach or choosing always the majority class (here that constitutes a baseline accuracy of~54~\%). 

We test the eleven classifiers using different approaches of feature selection: no/manual feature selection, SFS, and RFECV. 
In manual selection, we select subsets of features split by the physiological signals they were derived from. 
For each feature selection setting, we test none, min-max, and z-score normalization. 
We found min-max normalization works best with respect to our dataset and classifiers. 
Hence, for the analysis we focus only on min-max normalization. 
We train and validate each classifier twelve times to obtain a mean accuracy over all participants, see Tab.~\ref{tab:acc_all_classifiers}. 
Due to limited space, we only show results for subsets of PPG features, and PPG and EDA features because they are best in accuracy. 
We employ SHAP~\cite{Lundberg2017} to find the most descriptive features of the classifiers. 

\begin{table}[h!]
\caption{Overview of the mean accuracies over all participants obtained by our selected state-of-the-art classifiers using various feature selection algorithm and min-max normalization for all participants. The ``*" marked classifiers can have varying results due to their implementation.}
\centering
\begin{tabular}{|c|c|c|c|c|c|} 
\hline
Classifier & None & SFS & RFECV & PPG & PPG+EDA \\ [0.5ex] 
\hline\hline
SVC & \textbf{0.7917} & \textbf{0.7708} & 0.7083 & 0.6875 & \textbf{0.7917} \\
\hline
DTC* & 0.5417 & 0.5625 & 0.5625 & 0.5833 & 0.5625 \\
\hline
KNN & 0.75 & 0.6458 & N.A. & 0.7292 & 0.6458 \\
\hline
GNB & 0.6042 & 0.5833 & N.A. & 0.6667 & 0.6042 \\
\hline
LR & 0.5625 & 0.6875 & 0.6667 & 0.6875 & 0.7083 \\
\hline
LDA & 0.7708 & 0.6458 & \textbf{0.75} & \textbf{0.7708} & \textbf{0.7917} \\
\hline
QDA & 0.625 & 0.5625 & N.A. & 0.6042 & 0.625 \\
\hline
RF* & 0.6875 & 0.5208 & 0.7083 & 0.6875 & 0.625\\
\hline
GB* & 0.6458 & 0.5417 & 0.625 & 0.6042 & 0.5833 \\
\hline
AB & 0.6458 & 0.5 & 0.5833 & 0.6458 & 0.6458 \\
\hline
XGB & 0.5625 & 0.4167 & 0.6458 & 0.6667 & 0.5833\\
\hline
\end{tabular}
\label{tab:acc_all_classifiers}
\end{table}

\subsection{None and manual feature selection}
The SVC reached the highest mean accuracy~(79~\%) for all features and together with the LDA classifier a similar accuracy with features derived from PPG and EDA. 
The worst accuracy is achieved using the decision tree classifier~(54~\%) and XGB~(56~\%). 
When inspecting the individual accuracies, we find that participant~4 behaves dissimilar to the rest of the participants. 
While the SVC obtains 75~\% to 100~\% accuracy on all other participants, it obtains 0~\% for participant~4. 
This pattern, of performing much worse than all the other participants, holds for all classifiers. 
Applying the SHAP analysis we see that \textit{variability of tonic EDA} is a meaningful feature for SVC, KNN, and LDA (the three best performing classifiers).
We observe that reducing the number of features, for example, by only using PPG and EDA features, can increase the classifiers performance. 

%
\subsection{Sequential feature selection (SFS)}
We allow the algorithm to select~12~out of the~24~features (default setting). 
The mean accuracy is slightly lower than with out the feature selection algorithm and the support vector classifier is performing best with a mean accuracy of 77~\%. 
Using the SHAP analysis on the support vector classifier combined with SFS, we find that \textit{breathing rate}, \textit{SCR Peaks Amplitude Mean}, and \textit{EDA Sympathetic} are selected most often as a feature (11~out of~12~times) with \textit{breathing rate} having the highest mean($|$SHAP~value$|$) (for simplicity, SHAP~value). 
For the LR classifier (second best performing classifier in this setting) \textit{area of ellipse of sd1 and sd2} (\textit{s}), \textit{interbeat interval} (\textit{ibi}), \textit{portion of pairs of successive NNs that differ by more than 20~ms} (\textit{pNN20}), and \textit{EDA Sympathetic} are chosen most often (10~out of~12~times), with \textit{s} having the highest SHAP~value. 
The feature that was used most often with all classifiers is \textit{pNN20} with (avg. 9.09~out of~12~times). 
The least often picked feature is \textit{number of occurrences of SCR} (3.45~out of~12~times). 
The most descriptive feature over all classifiers is \textit{s} with a SHAP~value of 0.5859. 
The least descriptive feature is \textit{PSD power} of the temperature data SHAP~value~=~0.030). 
Overall, there is no feature that is dominant throughout all classifiers and participants.

\subsection{Recursive feature elimination with cross-validation (RFECV)}
This algorithm chooses the number of used features by itself using five-fold-cross-validation (default setting). 
The accuracies are similar to the SFS algorithm and slightly lower than not using a feature selection algorithm.  
The LDA classifier is performing best with a mean accuracy of 75~\%. 
The worst performing classifier is the decision tree classifier~(56~\%).
The k-nearest neighbor classifier, Gaussian Naive Bayes, and quadratic discriminant analysis cannot be used in combination with RFECV because they do not implement any feature importance measure. 
Hence, we do not consider them for the RFECV setting. 

As the algorithm determines itself how many features to select, we observe that \textit{s} is the most often selected feature (avg. 9.75 out of 12 times) followed by \textit{heart beats per minute} (\textit{bpm}) (avg. 9.375 out of 12 times), and \textit{standard deviation of NN intervals} (\textit{sdNN}) (avg. 8.5 out of 12 times). 
The least often selected feature is \textit{portion of pairs of successive NNs that differ by more than 50~ms} (avg. 3.5 out of 12 times) followed by \textit{root mean squared difference of RR intervals} (avg. 3.875 out of 12 times) (\textit{rmssd}), and \textit{standard deviation of differences of RR intervals} (avg. 4 out of 12 times). 
The most descriptive feature using the SHAP analysis is \textit{rmssd} (outlier in LR classifier participant~6) followed by \textit{sd1}, and \textit{bpm}. 
The least descriptive feature is the \textit{breathing rate} followed by \textit{temperature gradient}, and \textit{SCR~Peaks~Amplitude~Mean}. 
Similar to the observation for the SFS algorithm, we do not observe any dominant features throughout all classifiers and participants.

\section{Discussion}
\label{sec:discussion}
We extracted 24~features from PPG, EDA, and temperature data collected during air traffic controller training sequences. 
Based on these features we estimated the air traffic controllers' perceived passage of time under various workload conditions using state-of-the-art machine learning classifiers. 
We extend our analysis by two feature selection algorithms, SFS and RFECV, because manually selecting feature subsets showed that reducing the used features can simplify the problem and increase performance. 
Last, we applied SHAP analysis to measure feature importance. 

Our results show that eight out of eleven classifiers are able to classify the perceived passage of time sufficiently enough (over 60~\% accuracy) for previously unseen participants. 
However, we noticed that not all participants provide qualitatively equivalent feedback. 
For example, participant~4 consistently achieves low classification accuracies across all classifiers, that is, participant~4 has a different physiological response to the applied stimulus compared to all other participants. 
Considering SVC, LDA, and KNN for participant~4, the features derived from the thermopile (\textit{temperature gradient} and \textit{PSD Power}) have a larger influence than the features derived from EDA (\textit{variability of tonic EDA}), which are meaningful features for all other participants.
This implies that we can increase the classifier accuracy by fine-tuning our classifiers to individual users, which could be done by means of online learning, for example, during ``adaptive configuration" of the \textit{ChronoPilot}-device. 

\subsubsection{None and manual feature selection}
The decision tree classifier has the worst performance because it uses only a small subset of all possible features to build its decision tree. 
Even though we use the best splitting strategy the dataset is too small so not all features are required to optimally split the training dataset. 
This explains the varying results between runs (also holds for the random forest and GB classifier, which are ensembles of decision or regression trees).  
The SVC, LDA, and kNN classifier performed best. 
For all of these classifiers \textit{variability of tonic skin conductance level} is a meaningful feature. 
This finding can be explained by the fact that EDA is a proxy for quantifying cognitive load~\cite{Vanneste2021} and the perceived passage of time labels correlate with the experimental setting, that is, with the induced mental load. 
Other worse performing classifiers rely less on EDA derived features causing less accuracy.
The other classifiers show no patterns in their use of biomarkers based on the SHAP analysis. 
Interestingly, all types of physiological data are incorporated for classifying. 
However, not all classifiers perform best when provided with all available information, indicating that feature selection is indeed a useful tool in particular combinations of scenarios and classifiers.


\subsubsection{Sequential feature selection}
Using the SFS, half the features are chosen in a greedy approach. 
Accuracies drop for all but the decision tree classifier and logistic regression classifier. 
We achieve a higher accuracy using the DTC because SFS forced it to use 12~features, which was not the case using no feature selection algorithm (previously 6.4~features selected in average). 
The SFS caused to disregard the second most descriptive feature \textit{sdNN} of the LR classifier and replace it with the previously hardly considered feature \textit{ibi}, which may explain its increased performance. 
Although the mean accuracy of SVC over all participants slightly decreased from 79~\% to 77~\%, the SFS was able to increase the minimal accuracy to 50~\% (reached by participant~7) from 0~\% (participant~4) using no feature selection algorithm. 
For participant~4 it achieves an accuracy of 75~\%. 
These results suggest that the SFS algorithm can increase robustness for training our models on diverse participants with the trade off of reduced maximum accuracy.

\subsubsection{Recursive feature selection with cross-validation}
Using the RFECV algorithm cross-validation is used to find the least important feature, which can be removed. 
The best performing classifier LDA used in average 17.6 features over all participants. 
The selected features show comparable SHAP~values compared to no feature selection algorithm. 
DTC and RF are able to slightly increase their accuracy compared to not applying a feature selection algorithm. 
The increase of mean accuracy in the decision tree classifier could be due to favored random feature selection. 
Due to the RFECV the random forest classifier slightly increased the importance of EDA based features, which could explain the better performance. 
The same holds for the slight performance gains of the XGB classifier. 

The largest increase in mean accuracy is obtained using the logistic regression classifier (approximately 10~\%). 
Based on the SHAP values, we can argue that again the EDA based features are valued higher alongside the temperature values.

\subsubsection{Limitations}
Considering the future application scenario of providing feedback to a~\textit{ChronoPilot}-device, we encounter several practical limitations for data collection. 
We require live-stream sensor data and a reliable communication protocol to reduce latency and to mitigate data loss. Also reducing data noise and removing artifacts need to work online.
Depending on the use case, users may be required to be mobile (e.g., our industrial production scenario~\cite{Botev2021}). 
Motion artifacts may corrupt the physiological data, such as PPG. 
Hence, we may limit ourselves to physiological data that can be collected even if users are mobile and in motion. 
One possible PPG replacement is measuring electrocardiography (ECG), which directly measures the heart electrical activity. 
ECG allows for extracting the same features as PPG  
and there are methods to remove artifacts automatically.
Using a thermopile as sensor also highly depends on the later use case, for example, considering again the industrial production scenario the user has to stand or walk around. 
One possible replacement could be measuring the skin and surrounding temperature. 
Skin temperature can often be measured together with EDA using the same device. 
Our results indicate that EDA is a strong descriptive feature and should definitely be used in further studies. 
Another interesting information source could be eye tracking, as it provides information about the user's focus~\cite{Balzarotti2021} and it could be integrated in a wearable \textit{ChronoPilot}-device, such as VR/AR headsets. 




\section{Conclusion}
\label{sec:conclusion}
We were able to show that it is feasible to estimate the perceived passage of time of air traffic controllers based on physiological data, such as PPG, EDA, and temperature measurements of a thermopile during helicopter training sequences of varying workload conditions with high accuracy up to 79~\%. 
EDA seems to provide the most descriptive features based on SHAP analysis. 
This finding reinforces our assumption that we can influence the perceived passage of time by varying mental workload given that EDA is considered a biomarker for mental workload and stress. 
In terms of our objective of building a \textit{ChronoPilot}-device, we have shown that we can use physiological data as feedback to close the control loop.

For our future work, larger datasets may improve the chances for generalizability.
This can be done in two ways: increasing the number of participants or the amount of data per participant. 
In addition, larger datasets would allow us to explore deep learning methods. 
The literature on exploiting physiological data to automatically detect a person's subjective time perception is scarce. To date, a~closed-loop approach for guided manipulation of time perception has not yet been attempted. 
Our results reported here encourage us in our effort of creating an automatic closed-loop approach for modulating subjective time perception in real-world scenarios as envisioned in the \textit{ChronoPilot} project. By exploiting the plasticity of subjective time perception, we can develop future technology that enhances productivity and well-being.

\addtolength{\textheight}{-12cm}   





\section*{ACKNOWLEDGMENT}

This work was supported by European Union’s Horizon 2020 FET research program under grant agreement No.~964464 (ChronoPilot). We thank Andreas Psarrakis for assistance in the data collection process of this study.


\bibliographystyle{IEEEtran}
\bibliography{IEEEabrv,root}

\end{document}